\documentclass[11pt]{article}

\usepackage[usenames]{color}
\usepackage{amssymb}
\usepackage{amsmath}
\newcommand{\half}{\frac{1}{2}}
\newcommand{\del}{\partial}
\newcommand{\ep}{\epsilon}
\newcommand{\bra}[1]{\langle #1 |}
\newcommand{\ket}[1]{| #1 \rangle}
\newcommand{\bracket}[2]{\langle #1| #2 \rangle}

\newcommand{\nn}{\nonumber}

\newcommand{\bea}{\begin{eqnarray}}
\newcommand{\eea}{\end{eqnarray}}
\newcommand{\ap}{\alpha}

\newcommand{\Sym}{{\rm Sym}}
\newcommand{\mV}{2i\{b_0^-,O\}}
\newcommand{\ws}{{\rm WS}}
\newcommand{\bulk}{{\rm bulk}}

\newcommand{\ev}[1]{\left \langle #1 \right \rangle}

\begin{document}

\begin{titlepage}
\title{
\hfill\parbox{4cm}
{\normalsize{\tt hep-th/0610171}}\\
\vspace{1cm}
{\bf Reformulation of Boundary String Field Theory\\ in terms of Boundary State}
}
\author{
Shunsuke {\sc Teraguchi}
\thanks{{\tt teraguch@phys.sinica.edu.tw}}
\\[5pt]
{\it Institute of Physics, Academia Sinica, Taipei 11529, Taiwan}
}
\date{\normalsize October, 2006}
\maketitle
\thispagestyle{empty}

\begin{abstract}
We reformulate bosonic boundary string field theory in terms of boundary state.
In our formulation, we can formally perform the integration of target space equations of motion for arbitrary field configurations without assuming decoupling of matter and ghost.
Thus, we obtain the general form of the action of bosonic boundary string field theory.
This formulation may help us to understand possible interactions between boundary string field theory and the closed string sector.
\end{abstract}

\end{titlepage}

\section{Introduction}
Boundary string field theory (BSFT) \cite{Witten:1992qy, Witten:1992cr} is one of formulations of open string field theory.
In the analyses of off-shell open string tachyon dynamics\footnote{
See the review \cite{Sen:2004nf}, for example.}, BSFT has played important roles.
BSFT is applied to derive the exact form of tachyon potential \cite{Gerasimov:2000zp,Kutasov:2000qp} and the effective action derived from this string field theory has been used in a large amount of papers.

Though the action of BSFT is often identified simply with the partition function of the two-dimensional sigma model with boundary perturbation (and this is true for some specific cases), the original definition of this theory is rather complicated and abstract. In the case of bosonic string theory \cite{Witten:1992qy, Witten:1992cr}, the action $S$ of BSFT is defined by the following equation,
\bea
dS=\half\int_0^{2\pi}d\theta d\theta^\prime\ev{d{\cal O}(\theta)\{Q_B,{\cal O}(\theta^\prime)\}}_\lambda,
\label{originaldefofS}
\eea
where the unnormalized expectation value $\ev{\cdots}_\lambda$ is evaluated using a two-dimensional field theory on disk
whose boundary deformation from a conformal field theory (CFT) is specified by world-sheet couplings $\lambda$'s.
While the operator ${\cal O}(\theta)$ is defined on the boundary which is parametrized by $\theta$, $Q_B$ is the bulk BRST operator defined on the bulk CFT.
It is proved \cite{Witten:1992qy,Shatashvili:1993ps} that the right-hand side of the above equation is closed.
Then, at least locally, we can define the action $S$ through the above equation.
The action $S$, defined in this way, is proved to have a gauge symmetry based on the Batalin-Vilkovisky (BV) formalism.
The boundary world-sheet couplings $\lambda$'s are interpreted as string fields.
Unfortunately, explicit evaluation of the above quantity is possible only for very limited field configurations where the two-dimensional field theory is solvable.
However, in the case where matter and ghost are decoupled, one can formally perform the integration of the above equation
and it has been conjectured that the action $S$ takes the following form \cite{Witten:1992cr, Shatashvili:1993ps, Shatashvili:1993kk}:
\bea
S=-\beta^i(\lambda)\frac{\del}{\del \lambda^i}Z(\lambda)+Z(\lambda),
\eea
where $Z(\lambda)$ is the partition function of the two-dimensional sigma model and
$\beta^i(\lambda)$ is the world-sheet $\beta$-function with respect to the coupling $\lambda^i$.
On shell, where the $\beta$-functions vanish, the action reduces to the partition function itself.
For superstring theory, the BSFT formulation leads us to rather simple action \cite{Marino:2001qc, Niarchos:2001si},
\bea
S=Z(\lambda),
\eea
under the assumption of decoupling of matter and ghost.

Because BSFT is closely related to the partition function as we have seen above, 
one might suspect that the boundary state formalism is quite suitable to rewrite BSFT.
Using the boundary state $\bra{B}$, the partition function of the system can be written as
\bea
Z=\bracket{B}{0},
\label{innerproduct_pf}
\eea
where $\ket{0}$ is the closed string vacuum of first quantized string.
This inner product would give us the action of super BSFT or on-shell bosonic BSFT when there is no mixing between matter and ghost.
Because boundary states are states in closed string Hilbert space, this kind of reformulation of BSFT should allow us to
discuss interactions with its closed string sector.
However, these expressions rely on the assumption of decoupling of matter and ghost
and this assumption may spoil gauge invariant arguments.
We are also limited to on-shell case for bosonic string.
In this paper, we shall solve this problem.
Namely, we construct the basic ingredients of BV formulation in terms of closed string states
and show that they satisfy the same conditions as the original ones did.
Then we can rewrite the definition of the BSFT action $S$ corresponds to eq.(\ref{originaldefofS})
in terms of closed string states.
Surprisingly, we can formally perform the integration of this equation
without assuming the decoupling of matter and ghost in this formalism.

The contents of this paper is as follows.
The next section is devoted to a brief review of the bosonic BSFT.
In section 3, we propose how to construct basic ingredients of BV-formalism in terms of closed string language.
Using these ingredients, we can redefine the action of BSFT as in the original proposal of BSFT.
In the following section,
we formally evaluate the action in our formulation and obtain the form of the action itself.
This form (\ref{ourBSFTaction}) of the BSFT action represented in closed string states is our main result of this paper.
We verify the validity of this expression by calculating the well-known tachyon action in section 5.
Finally, we conclude this paper with discussions on several future directions.
Our convention in the text is summarized in the appendix.

\section{A Short Review of Bosonic BSFT}
In this section, we shall briefly review the construction of bosonic BSFT.
The action of BSFT is formally considered as a functional on the space of two-dimensional field theories which are parametrized by boundary deformations.
A point on the space of two-dimensional field theories should be specified by a set of world-sheet couplings $\lambda$'s.
The corresponding world-sheet action is given by
\bea
S_\ws=S_\bulk+\int_0^{2\pi}\frac{d\theta}{2\pi}{\cal V}(\lambda),
\label{wsaction}
\eea
where $S_\bulk$ is a bulk CFT action for matter and ghost which defines closed string background of this system.
Because BSFT is a string field theory for open string sector, this bulk CFT has to be fixed. 
The boundary operator ${\cal V}$ may be expanded in terms of boundary world-sheet couplings $\lambda$'s as
\bea
{\cal V}(\lambda)=\sum_{i}\lambda^i{\cal V}_i(\theta),
\eea
where ${\cal V}_i$ is a basis of the boundary operators with ghost number zero.
The following expansion might be more familiar from the point of view of the usual sigma model approach:
\bea
{\cal V}=T(X(\theta))+\del_\theta X^\mu A_\mu(X(\theta))+\cdots.
\eea
Thus, the world-sheet couplings $\lambda$'s are corresponds to the degrees of freedom of open string fields.
More precisely, we represent the boundary deformation $\cal V$ through ghost number one operator $\cal O$, whose relation to $\cal V$ is given by
\bea
{\cal V}=b_{-1}^{\rm BSFT}{\cal O}.
\label{originalantighostoperation}
\eea
The anti-ghost operator $b_{-1}^{\rm BSFT}=\oint b(v)$ is an analogue of the anti-ghost operator $b_{-1}$ in cubic open string field theory \cite{Witten:1985cc} and defined by the integration of an operator valued closed one form,
\bea
b(v)=v^ib_{ij}\ep^j_kd\sigma^k,
\eea
where $v^i$ and $\ep^j_k$ are the Killing vector and the complex structure on the disk, respectively.
The integration contour is chosen along the boundary of the disk and effectively enclose the boundary operator.
As they are in cubic open string field theory, these operators $\cal O$'s with ghost number one are
 considered as basic objects of this string field theory.

As is well-known, the guiding principle to construct a string field theory is the stringy gauge invariance.
In the construction of BSFT, the requirement of gauge invariance is fulfilled with the help of BV formalism.
BV formalism ensures that, if the action $S$ satisfy the BV master equation,
\bea
\{S,S\}=0,
\eea 
this action automatically has gauge invariance.
The anti-bracket appearing in the above BV master equation is defined with a non-degenerate closed two-form $\omega$ with ghost number $-1$,
\bea
\{A,B\}\equiv\frac{\del_r A}{\del \lambda^K}\omega^{KL}\frac{\del_l B}{\del \lambda^L},
\eea
where $\frac{\del_r}{\del \lambda^K}$ ($\frac{\del_l}{\del \lambda^K}$) is the right (left) derivative, respectively.
Here we allowed $\lambda$'s to include anti-fields whose statistics are opposite of corresponding fields.
One can construct an action functional $S$ which satisfies BV master equation by the following equation:
\bea
dS=i_V\omega,
\label{genDefOfS}
\eea
where $V$ is a nilpotent vector which generates a symmetry of two-form $\omega$.
The fact that $V$ generates a symmetry of two-form $\omega$,
$(di_V+i_Vd)\omega=0$,
and the closedness of $\omega$, $d\omega=0$, 
ensures that eq.(\ref{genDefOfS}) is integrable at least locally.
The nilpotency of $V$, $V^2=0$, indicates the functional $S$ defined in eq.(\ref{genDefOfS})
satisfies the BV master equation.

In bosonic BSFT, the nilpotent vector $V$ is defined through the action of the bulk BRST operator $Q_B$ on boundary operators $\cal O$,
\bea
\delta_V {\cal O}(\theta)=\{Q_B,{\cal O}(\theta)\}.
\label{odefofV}
\eea
The nilpotency of the vector $V$ immediately follows from the nilpotency of $Q_B$.
The two-form $\omega$ is given by
\bea
\omega_{IJ} = \half \int_0^{2\pi}\frac{d\theta d\theta^\prime}{(2\pi)^2}\ev{\delta {\cal O}_I(\theta)\delta {\cal O}_J(\theta^\prime)}_\lambda,
\label{originalomegadef}
\eea
where the correlator should be evaluated in the world-sheet theory (\ref{wsaction}).
It is proved \cite{Witten:1992qy, Shatashvili:1993ps} that the two-form $\omega$ defined in (\ref{originalomegadef}) satisfies the following requirements,
\bea
d\omega=0, \ d(i_V\omega)=0.
\eea
Then, the action $S$ of bosonic BSFT is given by
\bea
dS=\half\int_0^{2\pi}\frac{d\theta d\theta^\prime}{(2\pi)^2}\ev{d{\cal O}(\theta)\{Q_B,{\cal O}(\theta^\prime)\}}_\lambda.
\label{originaldefofbsft}
\eea
In order to obtain explicit form of this action, we must solve the two-dimensional field theory which corresponds to
the boundary deformation specified by string field $\lambda$.
In general, it is very difficult to solve such an interacting field theory.
Here we only review the well-known tachyon configurations \cite{Witten:1992cr} where the system is still free.
We choose the following configurations of the string field,
\bea
{\cal O}(\theta)=c^{\theta}(\theta)\left(a+\frac{1}{4}\sum_i u_i(X^i)^2(\theta)\right),
\label{freestringfield}
\eea
which gives us a purely matter boundary deformation,
\bea
{\cal V}(\theta)=a+\frac{1}{4}\sum_i u_i(X^i)^2(\theta).
\label{freedeformation}
\eea
In our convention, the anti-ghost operator
is given by
\bea
b_{-1}^{BSFT}
%=\int\frac{dw}{2\pi i}b(w)-\int\frac{d\bar w}{2\pi i} \tilde b(\bar w)
=-i\oint\frac{dz}{2\pi i}zb(z)-i\oint\frac{d\bar z}{2\pi i}\bar z \tilde b(\bar z),
\label{BSFTantighost-1}
\eea
where $z=e^{-i\theta+\tau}$ is the complex coordinates whose unit circle, $|z|^2=1$, specifies the disk. 
Note that, though the anti-ghost operator $b_{-1}^{BSFT}$ (\ref{BSFTantighost-1}) behaves as $b_{-1}$ of open string theory on operators at the boundary, it acts as $b_0-\tilde b_0$ on operators at the origin, as pointed out in the original paper
\cite{Witten:1992qy}.
In \cite{Witten:1992cr}, the two-dimensional field theory with boundary deformation (\ref{freedeformation}) was explicitly solved. 
After explicit calculation, it was found that eq.(\ref{originaldefofbsft}) takes the following form:
\bea
dS(a,u)=d\left(\sum_i \left(u_i-u_i\frac{\del}{\del u_i}\right)Z(a,u)
+(1+a)Z(a,u)\right).
\label{quadratic's ds}
\eea
Here, the partition function $Z$ is given by
\bea
Z(a,u)=e^{-a}\prod_i\sqrt{u_i}e^{\gamma u_i} \Gamma(u_i).
\label{original partition function}
\eea
Finally, after integration of the eq.(\ref{quadratic's ds}), the BSFT action corresponds to the 
string field configuration (\ref{freestringfield}) is obtained as
\bea
S(a,u)=\left(- \sum_i u_i\frac{\del}{\del u_j}-\left(a+\sum_i u_i\right)\frac{\del}{\del a}+1\right)Z(a,u).
\label{efofaction}
\eea

Later on, it was discussed \cite{Witten:1992cr, Shatashvili:1993ps, Shatashvili:1993kk} that, when matter and ghost is completely decoupled, namely, when $\cal V$ does not contain ghost operators, we can formally evaluate eq.(\ref{originaldefofbsft}) as
\bea
S=\frac{i}{2}\left(\ev{[L_{-1},\int d\theta e^{i\theta}{\cal V}(\theta)]}_\lambda+{\rm c.c}\right)
-\ev{\int d\theta{\cal V}(\theta)}_\lambda+Z(\lambda).
\label{decoupled action}
\eea
Furthermore, it was conjectured that bosonic BSFT takes the following expression:
\bea
S=-\beta_i(\lambda)\frac{\del}{\del \lambda^i}Z(\lambda)+Z(\lambda).
\eea

\section{Rewriting bosonic boundary string field theory in terms of boundary state}
One may notice that the coordinate system $z$ and $\bar z$ in the previous section,
is suitable to define closed string creation-annihilation oscillators in the bulk, rather than open string oscillators on the boundary. Furthermore, it is well-known that a partition function of a boundary deformed theory can be given by a boundary state, $\ket{B}$,
\bea
Z=\bracket{B}{0}.
\eea
Then it is rather natural to suspect that BSFT prefers to be formulated in terms of closed string Hilbert space.
Such reformulation may help
to investigate interactions with the closed string sector \cite{open+closed}.
However, many of such discussions rely on the case of on-shell BSFT or super BSFT without mixing matter and ghost,
where we can simply write the action as
\bea
S=Z.
\eea
Unfortunately, it is not clear what kind of modifications we should have for off-shell bosonic string theory case,
whose closed string field theories \cite{closedSFT} are relatively well understood.
In order to solve this difficulty, in this section, we reformulate bosonic BSFT in terms of boundary state based on BV formalism,
by replacing the basic ingredients of the previous section by corresponding ones represented in closed string Hilbert space.

We propose the following definitions of the nilpotent vector $V$ and the closed two-form $\omega$,
\bea
\delta_V {\cal O}(\theta)&=&\{Q_B,{\cal O}(\theta)\}, \label{defofV}\\
\omega&=&\half\bra{N}\Sym[e^{\mV};O_I, O_J]\ket{0}d\lambda^I\wedge d\lambda^J,
%%\omega_{IJ}=\bra{N}\Sym[e^{\mV};\delta O_I,\delta  O_J]\ket{0}
\label{defofomega}
\eea
where we have used a simplified notation for boundary operators,
\bea
O=\int_0^{2\pi}\frac{d\theta}{2\pi} {\cal O}(\theta).
\eea
$\bra{N}$ is the Neumann boundary state and the symbol $\Sym[\cdots]$ will be defined below.

Though, eq.(\ref{defofV}) looks completely same as before, eq.(\ref{odefofV}), there are two conceptual differences.
First, we are assuming that ${\cal O}(\theta)$ is written in terms of closed string oscillators\footnote{Note that, in order to define these operators ${\cal O}(\theta)$,
we have to take a limit where these operators approach the boundary.
This process requires suitable regularization (or normal-ordering).}.
Second, here the BRST operator is also written in terms of closed string oscillators:
\bea
Q_B&=&\frac{1}{2\pi i}\oint(dz j_B-d\bar z\tilde j_B)\nn\\
&=&c_0^+(L_0+\tilde L_0-2)+\half c^-_0(L_0-\tilde L_0)+(M+\tilde M)b_0^++2(M-\tilde M)b_0^-+Q_B^\prime,
\label{defofBRSTcharge}
\eea
where $L_0$ and $\tilde L_0$ are total Virasoro operators and
\bea
&&M=-\sum_{n=1}^\infty nc_{-n}c_n,\;\tilde M=-\sum_{n=1}^\infty n\tilde c_{-n}\tilde c_n,\\
&&Q_B^\prime=\sum_{n\neq0}\left(c_{-n}L^{\rm m}_{n}+\tilde c_{-n}\tilde L^{\rm m}_{n}\right)
+\sum_{n,m,n+m\neq 0}\frac{m-n}{2}\left(c_mc_nb_{-m-n}+\tilde c_m \tilde c_n \tilde b_{-m-n}\right).
\eea
In the original definition of BSFT, it is argued \cite{Shatashvili:1993ps} that the action of BRST operator can be affected
by boundary deformations when the counter approach to the boundary. Then,  in order to keep the integrability of the RHS of eq.(\ref{originaldefofbsft}), a suitable regularization which satisfies some constraint should be chosen.
In our definition, the action of BRST operator is defined regardless of boundary deformations.
On the other hand, such a effect of the boundary is encoded in the definition of closed two-form $\omega$.

In order to define the closed two-form $\omega$, we have introduced a symbol $\Sym[\cdots]$,
which represents parameter integrations with operator insertions:
\bea
&&\Sym[e^{-V};O_1,O_2,\cdots,O_n]\nn\\
&=&\int_0^1dt_1 \int_{t_1}^1dt_2 \cdots \int_{t_{n-1}}^1 dt_n  e^{-t_1V}O_1
e^{-(t_2-t_1)V}O_2\cdots O_n e^{-(1-t_n)V}
\pm({\rm perms}).
\eea
The signs in front of the permutation terms come from the fermionic property of operators
and also from the anti-symmetric property of differential forms.
Note that these parameter integrations are compatible with the deformation of the operator $V$ in the following sense:
\bea
\delta(e^{-V})=\Sym[e^{-V},-\delta V].
\eea
In eq.(\ref{defofomega}), we constructed two-form $\omega$ by inserting two operators on the boundary state,
\bea
\bra{B}\equiv\bra{N}\exp\left(\mV\right)
\eea
which is not necessary on-shell but formally satisfy the boundary condition specified by the boundary deformation.
We have rewritten the operation (\ref{originalantighostoperation}) using the anti-commutator with $2b_0^-=b_0-\tilde b_0$.

If these ingredients enjoy the following properties,
\bea
V^2=0:&\;\rm{nilpotency},\label{nilpotency}\\
d\omega=0:&\;\rm{closedness},\\
d(i_V\omega)=0:&\;V\rm{-invariance},\label{V-invariance}
\eea
we can construct a gauge invariant action $S$ as before.
The nilpotency of vector $V$ directly follows from the nilpotency of $Q_B$ again.
We can easily check the closedness of the $\omega_{IJ}$:
\bea
\frac{d}{d\lambda^K}\omega_{IJ}+({\rm perms})
&=&2i\bra{N}\Sym[e^{\mV};\{b_0^-,O_K\},O_I,O_J]\ket{0}+({\rm perms})\nn\\
&=&-2i\bra{N}\Sym[e^{\mV};O_K,O_I,O_J]b_0^-\ket{0}+({\rm perms})=0.
\eea
The second equality comes from the nilpotency of $b_0^-$.
The invariance of $\omega$ under the transformation generated by $V$ is more involved.
From the above definitions, we have two contributions for $d(i_V\omega)$.
One comes from the derivative of the exponential,
\bea
2i\bra{N}\Sym[e^{\mV};\{b_0^-,dO\},dO,\{Q_B,O\}]\ket{0},
\label{contributionA}
\eea
the other comes from the derivative of the inserted operator $O$,
\bea
-\bra{N}\Sym[e^{\mV};dO,\{Q_B,dO\}]\ket{0}.
\label{contributionB}
\eea
The first contribution (\ref{contributionA}) is evaluated as
\bea
-i\bra{N}\Sym[e^{\mV};dO,dO,[b_0^-,\{Q_B,O\}]]\ket{0},
\eea
and the second one (\ref{contributionB}) gives
\bea
-i\bra{N}\Sym[e^{\mV};dO,dO,[Q_B,\{b_0^-,O\}]]\ket{0}.
\eea
These two combine into
\bea
&&-i\bra{N}\Sym[e^{\mV};dO,dO,[\{b_0^-,Q_B\},O]]\ket{0}\nn\\
&=&-\frac{i}{2}\bra{N}\Sym[e^{\mV};dO,dO,[L_0-\bar{L}_0,O]]\ket{0}.
\label{divwlast}
\eea
We must require that this term should vanish.
Though the operator $L_0-\bar{L}_0$ generates 
the rotation of the disk for the coordinate $z$, it is more straightforward to express this term using the coordinate
$w=i\log z=\theta+i\tau$. The boundary operator ${\cal O}(\theta)$ does not have any coefficient which depends on $\theta$
if this operator is expressed in this coordinate.
In this coordinate, $i[L_0-\bar{L}_0,{\cal O}(\theta)]$ takes the form of
\bea
\left(\int\frac{dw^\prime}{2\pi i}T(w^\prime)+\int\frac{d\bar w^\prime}{2\pi i}\tilde{T}(\bar w^\prime)\right)
{\cal O}(w,\bar w)=\del_\theta{\cal O}(w,\bar w).
\eea
The integration of this operator over $\theta$ gives zero and hence eq.(\ref{divwlast}) vanishes.

Note that, in the original formulation, it is pointed out in \cite{Shatashvili:1993ps} that
such a total derivative in the expectation value does not necessary vanish due to the boundary effect.
Furthermore, the action of $Q_B$ may also be affected by the boundary deformation and
$dQ_B$ should be included in the above evaluation.
Then, it is concluded that we
must choose suitable regularization where the following condition is satisfied:
\bea
-\ev{dO\{dQ,O\}}_\lambda+\half\ev{(dO)^2[L_0,O]}=0.
\eea
On the other hand, in our formulation,
we do not have any contribution from the derivative of $Q_B$ because we defined it in terms of closed string oscillators
by eq.(\ref{defofBRSTcharge}).
$[L_0-\bar{L}_0,O]$ is also simply evaluated regardless of the boundary deformation and vanishes
as far as we choose a suitable regularization when we define a limit where the bulk operator approaches the boundary\footnote{
In other words, we are not allowed to include deformations which do not satisfy the requirement $[L_0-\bar{L}_0,O]=0$ as open string degrees of freedom. 
}.
Thus, the above quantities are separately zero in our formulation.

We have proved that the three conditions (\ref{nilpotency})-(\ref{V-invariance}) are satisfied
based on our new definition of the nilpotent vector $V$ (\ref{defofV}) and the closed two-form $\omega$ (\ref{defofomega}).
Then, in terms of closed string Hilbert space, the action of BSFT can be defined as
\bea
dS=\bra{N}\Sym[e^{\mV};dO,[Q_B,O]\ket{0}.
\label{defofS}
\eea
The gauge invariance of this action is automatically guaranteed by BV-formalism.

In the above evaluation, we have used the following properties:
\bea
\bra{N}b_0^-=\bra{N}Q_B=b_0^-\ket{0}=Q_B\ket{0}=0.
\eea
Unfortunately, these conditions explicitly rely on the open string background which is specified
by the Neumann boundary state $\bra{N}$.
However, we can repeat the same argument based on another open string background $\bra{N^\prime}$
as far as this state satisfies above properties and construct a BSFT action based on
the perturbation from the background $\bra{N^\prime}$.
Furthermore,
by properly choosing a closed string vacuum, BRST operator and boundary state, it would be also 
possible to construct BSFT in a nontrivial closed string background.

\section{Formal evaluation of the action}
In the previous section, we repeated a similar argument discussed in the original proposal of BSFT
\cite{Witten:1992qy, Witten:1992cr, Shatashvili:1993ps, Shatashvili:1993kk} and obtained equations of motion of BSFT in terms of closed string states and oscillators. 
In this section, we shall further evaluate our new definition of the BSFT action (\ref{defofS}).
Surprisingly, without any assumption for the boundary deformation $\cal O$, we can perform the integration of eq.(\ref{defofS}) and find a formal expression of the action $S$ itself.

First we pull $b^-_0$ out from the expression (\ref{defofS}):
\bea
&&\bra{N}\Sym[e^{\mV};dO,\{Q_B,O\}]\ket{0}\nn\\
%&=&\bra{N}\Sym[e^{\mV};dO,\{Q_B,O\}]b_0^-c_0^-\ket{0}\nn\\
%&=&\bra{N}\Sym[e^{\mV};dO,[\{Q_B,O\},b_0^-]]c_0^-\ket{0}
%+\bra{N}\Sym[e^{\mV};\{b_0^-,dO\},\{Q_B,O\}]c_0^-\ket{0}\nn\\
&=&\bra{N}\Sym[e^{\mV};dO,[Q_B,\{b_0^-,O\}]]c_0^-\ket{0}
-\bra{N}\Sym[e^{\mV};dO,[\{b^-_0,Q_B\},O]]c_0^-\ket{0}\nn\\
&&+\bra{N}\Sym[e^{\mV};\{b_0^-,dO\},\{Q_B,O\}]c_0^-\ket{0}.
\label{ds_ex1}
\eea
The second term in the second line vanishes due to the rotational symmetry as before.
The first term consists of the following two terms.
\bea
\frac{i}{2}\bra{N}\Sym[e^{\mV};dO]Q_Bc_0^-\ket{0}-
\frac{i}{2}\bra{N}\Sym[e^{\mV};\{Q_B,dO\}]c_0^-\ket{0}.
\label{ds_ex2}
\eea
The second term of the above equation and the last term in (\ref{ds_ex1}) become a total derivative,
\bea
-\frac{i}{2} d\bra{N}\Sym[e^{\mV};\{Q_B,O\}]c_0^-\ket{0}.
\label{ds_ex3}
\eea
The first term of (\ref{ds_ex2}) can be evaluated by pulling $b^-_0$ out again,
\bea
\frac{i}{2}\bra{N}\Sym[e^{\mV};dO]b^-_0c^-_0Q_Bc_0^-\ket{0}
=\frac{1}{4}d\bra{N}e^{\mV}c^-_0Q_Bc_0^-\ket{0},
\eea
where we have used the fact that $[b^-_0,\{Q_B,c^-_0\}]$ vanishes.
As a result, we have evaluated eq.(\ref{defofS}) as a total derivative
\bea
dS=d\left(
\frac{1}{4}\bra{N}e^{\mV}c^-_0Q_Bc_0^-\ket{0}-\frac{i}{2} \bra{N}\Sym[e^{\mV};\{Q_B,O\}]c_0^-\ket{0}
\right).
\eea
Thus, we have obtained the action of BSFT itself as
\bea
S=\frac{1}{4}\bra{N}e^{\mV}c^-_0Q_Bc_0^-\ket{0}-\frac{i}{2} \bra{N}\Sym[e^{\mV};\{Q_B,O\}]c_0^-\ket{0}.
\label{ourBSFTaction}
\eea
This general form of BSFT action is the our main result in this paper.
Though, BRST operator, $Q_B$, appears in the first term, only ghost parts of it survives due to two $c_0^-$
and it is further evaluated as
\bea
\frac{1}{4}\bra{N}e^{\mV}c^-_0Q_Bc_0^-\ket{0}
%=\frac{1}{2}\bra{N}e^{\mV}c^-_0(M-\tilde M)\ket{0}
%=-\frac{1}{2}\bra{N}e^{\mV}c^-_0(c_{-1}c_1-\tilde c_{-1}\tilde c_1)\ket{0}=
%-\bra{N}e^{\mV}c^-_0c_1\tilde c_1\ket{0}
=Z,
\eea
if we assume that $\bra{N}e^{\mV}(c_{\pm 1}+\tilde c_{\mp 1})=0$ still holds\footnote{
For example, this is the case if matter and ghost are decoupled.}.
Thus, the first term reduces to partition function $Z$.
On the other hand, the second term vanishes if the deformation is on-shell.
This term represents the correction from the naive expectation of BSFT action, $S\sim Z.$
Note that
there is a gap between the on-shell condition for open string modes and the one for closed string modes originating from the choices of normal orderings.
In our formulation, an on-shell boundary operator is on-shell only
when it is represented with a suitable normal-ordering.
If another normal ordering is chosen, such an operator would be understood as a linear combinations of off-shell operators.

\section{An Example}
In this section, we reconsider the well-known example (\ref{efofaction}) of bosonic BSFT and
check whether our formula (\ref{ourBSFTaction}) correctly reproduces the same result.
Though we take the quadratic tachyon configuration (\ref{freedeformation}) as before\footnote{For simplicity, we omit space-time indices $i$ in the following discussion.},
in order to define this operator we must choose a suitable boundary normal-ordering for $X^2(\theta)$.
Because, in the boundary normal-ordering, the subtraction should be doubled compared to the usual bulk normal-ordering, we use the following prescription to define the operator $X^2(\theta)$:
\bea
X^2(\theta)&\equiv& \left(x-i\sum_{m=1}^\infty\frac{1}{m}(\ap_{-m}e^{-im\theta}+\tilde\ap_{-m}e^{im\theta})\right)\left(x-i\sum_{m=1}^\infty\frac{1}{m}(\ap_{-m}e^{-im\theta}+\tilde\ap_{-m}e^{im\theta})\right)\nn\\
&&+\left(i\sum_{m=1}^\infty\frac{1}{m}(\ap_{m}e^{im\theta}+\tilde\ap_{m}e^{-im\theta})\right)
\left(i\sum_{m=1}^\infty\frac{1}{m}(\ap_{m}e^{im\theta}+\tilde\ap_{m}e^{-im\theta})\right)\nn\\
&&+3\left(x-i\sum_{m=1}^\infty\frac{1}{m}(\ap_{-m}e^{-im\theta}+\tilde\ap_{-m}e^{im\theta})\right)
\left(i\sum_{m=1}^\infty\frac{1}{m}(\ap_{m}e^{im\theta}+\tilde\ap_{m}e^{-im\theta})\right)\nn\\
&&-\left(i\sum_{m=1}^\infty\frac{1}{m}(\ap_{m}e^{im\theta}+\tilde\ap_{m}e^{-im\theta})\right)
\left(x-i\sum_{m=1}^\infty\frac{1}{m}(\ap_{-m}e^{-im\theta}+\tilde\ap_{-m}e^{im\theta})\right)
\eea
Using this definition of $X^2(\theta)$, the boundary operator is given by
\bea
{\cal O}(\theta)=\left(\frac{i}{2}\sum_mc_me^{im\theta}-\frac{i}{2}\sum_m\tilde{c}_me^{-im\theta}\right)
\left(a+\frac{u}{4}X^2(\theta)\right).
\eea
This string field ${\cal O}(\theta)$ gives us the following boundary deformation,
\bea
{\cal V}(\theta)=-2i\{b_0^-, {\cal O}(\theta)\}=a+\frac{u}{4}X^2(\theta).
\eea
The first term of the BSFT action (\ref{ourBSFTaction}), namely, the partition function $Z$ can be evaluated using the following identity,
\bea
%\bra{0}\exp(-\frac{\ap_m\tilde\ap_m}{m})\exp\Bigg(-\frac{u}{2}\frac{1}{m^2}(\ap_m-\tilde\ap_{-m})(\ap_{-m}-\tilde\ap_{m})\Bigg)=\bra{0}\exp\left(-\ap_m\frac{m-u}{m(m+u)}\tilde{\ap}_m\right)(1+u/n)^{-1},
\bra{0}\exp(-a\tilde a)\exp\Bigg(-\frac{u}{2}(a-\tilde a^\dagger)(a^\dagger-\tilde a)\Bigg)=(1+u)^{-1}\bra{0}\exp\left(-a \frac{1-u}{1+u}\tilde{a}\right),
\eea
where $a,a^\dagger,\tilde a$ and $\tilde a^\dagger$ satisfy usual commutation relations,
$[a,a^\dagger]=[\tilde a,\tilde a^\dagger]=1$.
Thus, we have\footnote{
Similar derivation of boundary state and partition function can be found in \cite{Lee:2001ey}.},
\bea
\frac{1}{4}\bra{N}e^{\mV}c^-_0Q_Bc_0^-\ket{0}=Z
&=&2\sqrt{\frac{\pi}{u}}e^{-a}\prod_{n=1}^\infty\left(1+\frac{u}{n}\right)^{-1}e^\frac{u}{n}\nn\\
&=&2\sqrt{\pi}e^{-a}\sqrt{u}e^{\gamma u}\Gamma(u).
\eea
This partition function is exactly same as the one (\ref{original partition function}) calculated in the open string picture  (up to constant factor which we have not yet addressed in this paper).
For the calculation of the second term of the BSFT action (\ref{ourBSFTaction}), we only need the term 
linear to $c_1\tilde c_1$ in the commutator between $Q_B$ and ${\cal O}(\theta)$
(up to ghost boundary condition).
One of such terms comes from the commutator between  $Q_B$ and the ghost operator, and it is linear to the boundary deformation ${\cal V}$ itself,
\bea
-2ic_1\tilde c_1 {\cal V}(\theta).
\eea
The other term comes from the commutator between $c_{\pm1}L^{\rm m}_{\mp1}+\tilde c_{\pm1}\tilde L^{\rm m}_{\mp1}$ in $Q_B$ and the matter oscillators in ${\cal O}(\theta)$.
\bea
\int_0^{2\pi}\frac{d\theta}{2\pi}\{c_{\pm1}L^{\rm m}_{\mp1}+\tilde c_{\pm1}\tilde L^{\rm m}_{\mp1},{\cal O}(\theta)\}\sim-2iuc_1\tilde c_1.
\eea
Collecting these pieces, we have,
\bea
S[a,u]=\left(-u\frac{\del}{\del u}-(a+u)\frac{\del}{\del a}+1\right)Z[a,u].
\eea
Thus, we directly recovered the BSFT action (\ref{efofaction}) for quadratic tachyon configuration from our formula (\ref{ourBSFTaction}) which is expressed in terms of closed string Hilbert space.

\section{Conclusion and Discussion}
In this paper, we have reformulated the bosonic BSFT in terms of closed string Hilbert space.
Though the construction is almost straightforward, our formulation allows us to formally perform the integration.
Thus, we have obtained the action itself 
without any assumption for boundary operators, namely, without fixing gauge.
We have also checked that our formulation correctly reproduces the known result.

Though, our new formulation of BSFT looks working well within the scope of this paper, there is a subtle point related to normal-ordering prescription.
In general, commutators with BRST operator $Q_B$ depend on normal-ordering prescription.
For example, if a boundary operator is on-shell, it should commute with $Q_B$.
However, we have not yet understood how to take normal-ordering in general cases.
The normal-ordering which we took in this paper looks somehow ad hoc.
More transparent ways to understand this prescription are desirable.

There are several future directions.
The generalization to superstring may be straightforward.
In superstring case, it is well-known that the action of BSFT is simply given by the partition function when the matter and ghost are decoupled. However, it would be interesting to check whether it is also true for general operators.
Another direction might be to investigate the explicit form of the bosonic BSFT action.
In superstring case, we can simply take the partition function of the system as the action for physical states and there are several analyses based on the expansion of the partition function.
Our reformulation may help us to do similar analyses in bosonic case.
Furthermore, our final expression (\ref{ourBSFTaction}) is given in rather algebraic manner.
It may be possible to perform a brute-force evaluation of the action using computers.
The most interesting direction would be to investigate the possible interactions with the closed string sector.
Our action is already written in terms of closed string states,
it would be rather natural to expect this theory has proper couplings with the closed strings.\footnote{An incomplete list of such previous attempts can be found in \cite{open+closed,closed+open}.}
More ambitiously, we might be able to consider an open-closed string field theory where open string degrees of freedom are
realized as our BSFT action.
It has been pointed out \cite{Hashimoto:2004qp} that the BSFT based on two-dimensional field theory on a disk can not describe loop effects.
In this sense, BSFT may be understood just as a part of a more complete theory where closed string degrees of freedom are included and loop effects comes from closed string propagations. Especially, it would be very exciting if we could derive such a system from purely closed string field theory.

\section*{Acknowledgments}
We would like to thank K.~Furuuchi, K.~Hashimoto, H.~Hata, P.~Ho, I.~Kishimoto, S.~Moriyama, D.~Tomino and Y.~Yang for useful comments and discussions. 
S.T is supported by the Academia Sinica under grant
NSC95-2112-M-001-013.

\section*{Appendix}
\appendix
\section{Closed String Oscillators and Neumann Boundary State}
We take the convention of $\ap'=2$.\\
Mode expansion:
\bea
\del X^\mu(z)=-i\sum_{m=-\infty}^{\infty}\frac{\ap^\mu_m}{z^{m+1}}&,&
\bar\del X^\mu(\bar z)=-i\sum_{m=-\infty}^{\infty}\frac{\tilde\ap^\mu_m}{\bar z^{m+1}},\\
b(z)=\sum_{m=-\infty}^\infty\frac{b_m}{z^{m+2}}&,&\tilde b(\bar z)=\sum_{m=-\infty}^\infty\frac{\tilde b_m}{\bar z^{m+2}},\\
c(z)=\sum_{m=-\infty}^\infty\frac{c_m}{z^{m-1}}&,&\tilde c(\bar z)=\sum_{m=-\infty}^\infty\frac{\tilde c_m}{\bar z^{m-1}}.
\eea
Ghost zero-modes:
\bea
b_0^+\equiv b_0+\tilde b_0&,&b_0^-\equiv \half(b_0-\tilde b_0),\\
c_0^+\equiv \half(c_0+\tilde c_0)&,&c_0^-\equiv c_0-\tilde c_0,\\
\{b_0^\pm,c_0^\pm\}=1&,&\{b_0^\pm,c_0^\mp\}=0.
\eea
Normalization:
\bea
\bra{0}c_{-1}\tilde c_{-1}c_0^-c_0^+c_1\tilde c_1\ket{0}=1.
\eea
Neumann boundary state:
\bea
\bra{N}&=&\bra{0}c_{-1}\tilde c_{-1}c_0^+\exp\left(-\sum_{n=1}^\infty \frac{1}{n}\ap_{n}^\mu\tilde{\ap}_{n,\mu}
-\sum_{n=1}^\infty(c_{n}\tilde b_{n}+\tilde c_{n} b_{n})\right),
\eea
which satisfies the following relations:
\bea
\bra{N}(\ap_{n}^\mu+\tilde \ap_{-n}^\mu)=\bra{N}(c_n+\tilde c_{-n})=\bra{N}(b_n-\tilde b_{-n})=0.
\eea

\end{document}